\documentstyle[11pt,emulateapj,apjfonts,epsf]{article}
 
\def\kms{\relax \ifmmode {\,\rm km\,s}^{-1}\else \,km\,s$^{-1}$\fi}

\def\farcs{\hbox{$.\!\!^{\prime\prime}$}}
\def\arcdeg{\hbox{$^\circ$}}
\def\arcmin{\hbox{$^\prime$}}
\def\arcsec{\hbox{$^{\prime\prime}$}}
\def\secd#1.#2{ #1\farcs#2 }               

\received{\today}
\accepted{}
\slugcomment{The Astrophysical Journal Supplement Series}
 
\lefthead{}
\righthead{X-ray emission from PNe}
 
\begin{document}

\title{ROSAT observations of X-ray emission from planetary nebulae}

\author{Mart\'{\i}n A. Guerrero, You-Hua Chu, and Robert A. Gruendl}
\affil{Department of Astronomy, University of Illinois at Urbana-Champaign,\\ 
1002 West Green Street, Urbana, IL 61801\\
E-mail: mar@astro.uiuc.edu, chu@astro.uiuc.edu, gruendl@astro.uiuc.edu}

\submitted{{\it Received 1999 December 23; accepted 2000 January}}

\begin{abstract}
 
We have searched the entire ROSAT archive for useful observations
to study X-ray emission from Galactic planetary nebulae (PNs).  
The search yields a sample of 63 PNs, which we call the ROSAT 
PN sample.  
About 20--25\% of this sample show X-ray emission; these include 
13 definite detections and three possible detections (at a 
2$\sigma$ level). 
All X-ray sources in these PNs are concentrated near the central 
stars. 
Only A\,30, BD+30\arcdeg3639, and NGC\,6543 are marginally resolved 
by the ROSAT instruments.

Three types of X-ray spectra are seen in PNs.  
Type 1 consists of only soft X-ray emission ($<$0.5 keV), peaks at 
0.1--0.2 keV, and can be fitted by blackbody models at temperatures 
$1-2\times10^5$ K.  
Type 2 consists of harder X-ray emission, peaks at $>0.5$ keV, and 
can be fitted by thin plasma emission models at temperatures of a 
few 10$^6$ K.  
Type 3 is a composite of a bright Type 1 component and a fainter Type 
2 component.

Unresolved soft sources with Type 1 spectra or the soft component 
of Type 3 spectra are most likely photospheric emission from the hot 
central stars. 
Absorption cross sections are large for these soft-energy 
photons; therefore, only large, tenuous, evolved PNs with hot 
central stars and small absorption column densities have been 
detected.

The origin of hard X-ray emission from PNs is uncertain.  
PNs with Type 2 spectra are small, dense, young nebulae with 
relatively cool ($\ll10^5$ K) central stars, while PNs with Type 
3 X-ray spectra are large, tenuous, evolved nebulae with hot 
central stars. 
The hard X-ray luminosities are also different between these two
types of PNs, indicating perhaps different origins of their 
hard X-ray emission.
Future Chandra and XMM observations with high spatial and 
spectral resolution will help to understand the origin of hard 
X-ray emission from PNs.

\end{abstract}
 
\keywords{planetary nebulae: general -- stars: AGB and post-AGB -- X-rays: 
stars}

\section{Introduction}

Planetary nebulae (PNs) are expected to emit weakly at X-ray
wavelengths.  X-ray emission from a PN was not detected until the 
mid 1980's, when de Korte et al.\ (1985) reported EXOSAT observations 
of NGC\,1360.
Soon afterward, data from the Einstein archive were used to discover 
X-ray emission from NGC\,246, NGC\,6853, NGC\,7293, and A\,33 (Tarafdar 
\& Apparao 1988).
Similarly, data from the EXOSAT archive were used to show X-ray emission
from four more PNs, NGC\,1535, NGC\,4361, NGC\,3587, and A\,36 (Apparao 
\& Tarafdar 1989). 
More recently, ROSAT observations have yielded additional detections: A\,12, 
BD+30{\arcdeg}3639, LoTr\,5, and NGC\,6543 (Kreysing et al.\ 1992); K\,1-27 
(Rauch, Koeppen, \& Werner 1994); K\,1-16 (Hoare et al.\ 1995); and A\,30 
(Chu \& Ho 1995).  
To date, a total of 16 PNs have been reported to emit X-rays.

All of the detections made prior to the ROSAT era were interpreted as soft
X-ray emission from the hot (100,000--200,000~K) central star. 
In the more recent ROSAT observations, {\it diffuse} X-ray emission from PNs 
has been reported (Kreysing et al.\ 1992 -- A\,12, LoTr\,5, NGC\,4361, 
NGC\,6543 and NGC\,6853; Chu \& Ho 1995 -- A\,30; Leahy, Kwok, \& Yin 1998 -- 
BD+30\arcdeg3639).  This diffuse X-ray emission, if confirmed, is likely to 
originate from shock-heated gas.

ROSAT has provided both high spatial resolution to study the 
distribution of X-ray emission and high sensitivity to enable the spectral 
analysis of X-ray emission from PNs. 
Conway \& Chu (1997) noted that X-ray emission of PNs can be divided into 
three different spectral types: 
(1) spectra with only soft ($<$0.4 keV) X-ray emission, for example, 
NGC\,6853 (Kreysing et al.\ 1992), NGC\,246, NGC\,1360, and K\,1-16 
(Hoare et al.\ 1995); 
(2) spectra that peak at greater than 0.5~keV, for example, BD+30\arcdeg3639 
(Arnaud, Borkowski, \& Harrington 1996) and NGC\,6543 (Kreysing et al.\ 1992); 
(3) spectra with both soft and hard components, for example, NGC\,7293 
(Leahy, Zhang, \& Kwok 1994).

Despite the apparent success detecting X-ray emission from PNs, some 
misidentifications and erroneous analyses have appeared in the literature. 
The misidentifications include A\,12, A\,33, and NGC\,1535, as pointed out 
by Hoare et al.\ (1995), Conway \& Chu (1997), and Chu, Gruendl, \& Conway 
(1998), respectively.
A detailed comparison between the X-ray and optical images reveals that 
the X-ray sources are located outside the optical boundary of these PNs. 
The diffuse-nature of the X-ray emission reported in LoTr\,5, NGC\,4361, 
and NGC\,6853 has been disputed.  
The apparently extended morphology of LoTr\,5 and NGC\,4361 (Kreysing et al.\ 
1992) are probably caused by the low signal-to-noise ratio available in the 
ROSAT All Sky Survey data, as a deep pointed ROSAT observation of LoTr\,5 
shows an unresolved source (Chu \& Ho 1995), and an EUVE observation of 
NGC\,4361 detects only a point source (Fruscione et al.\ 1995).  
The apparently extended morphology in NGC\,6853 is caused by an electronic 
ghost image for photon energies below 0.2 keV (Chu, Kwitter, \& Kaler 1993).

ROSAT observations of PNs have discovered $\sim50\%$ of the currently known
X-ray sources in PNs.  Furthermore, ROSAT has provided, for the first time,
spatial and spectral information of X-ray emission from PNs.  The 
large volume of ROSAT observations offers an excellent opportunity for
archival studies.  We have searched the ROSAT archive for observations 
that contain PNs within the central, unvignetted field of view.  We find 
useful ROSAT observations for 80 PNs.  Some of the observations were pointed
at the PNs in question, while others were serendipitous observations pointed 
at objects projected near a PN.  
Only a small fraction of these observations of PNs have been analyzed and 
reported previously. 

The ROSAT archive allows us to assemble the most complete set of 
X-ray observations of PNs possible. 
We have analyzed these observations to determine the spatial and spectral 
properties of X-ray emission for the PNs detected and the 3$\sigma$ 
upper limits for the non-detections.
We have examined how absorption affects the detectability of hard and soft
X-ray sources in PNs.  We have further correlated the observed X-ray 
properties with extinction, distance, stellar and nebular properties,
in order to interpret the X-ray observations of PNs.
The results not only help us understand the origin of X-ray emission from 
PNs, but also provide a roadmap for future observations of PNs using the 
up-coming X-ray missions, e.g., Chandra Advanced X-ray Astrophysics 
Facility and X-ray Multi-mirror Mission.

This paper reports our work on ROSAT observations of PNs.  
In Section 2 we describe the PN sample and present our analyses and results. 
Section 3 discusses the detectability of X-ray emission from PNs, absorption
effects on the survey results, correlation between X-ray spectra and physical
properties of the PNs, and physical nature of X-ray emission from PNs.
A summary is given in Section 4.

\section{X-ray Survey of Planetary Nebulae in the ROSAT Archive}

\subsection{ROSAT Observations of Planetary Nebulae}

ROSAT (R\"Oentgen SATellite) had two types of X-ray detectors onboard:
the Position Sensitive Proportional Counter (PSPC) and the
High Resolution Imager (HRI).  
The PSPC is sensitive to X-rays in the energy range of 0.1--2.4 keV.  
Its on-axis point spread function (PSF) has a FWHM $\sim25\arcsec$ and 
its spectral resolution is $\sim45\%$ at 1 keV. 
The spectral resolution below 0.28 keV can be augmented if observations 
are made both with and without a boron filter, as the boron filter 
blocks X-rays between 0.188 and 0.28 keV. 
The PSPC field of view is $\sim1\arcdeg$ in radius, but the PSF deteriorates 
rapidly and vignetting becomes significant beyond {20\arcmin} from the center 
(ROSAT Mission Description 1991). 
The HRI is sensitive to X-rays in the energy range of 0.1--2.0 keV, but has 
negligible spectral resolution. 
It has an in-flight, on-axis PSF of FWHM $\sim6\arcsec$ and a field of view 
of $38\arcmin\times38\arcmin$ (David et al.\ 1996). 
The soft X-ray response of these detectors makes ROSAT observations 
invaluable to study X-ray emission from PNs. 
ROSAT archival data can be obtained from the anonymous ftp site  
legacy.gsfc.nasa.gov, or downloaded from the web site 
http://heasarc.gsfc.nasa.gov/W3Browse. 
Both of these sites are supported by the High Energy Astrophysics Science 
Archive Research Center (HEASARC) of Goddard Space Flight Center, NASA.

To search for useful ROSAT observations of PNs, we used the list of 
Galactic PNs in the ``Strasbourg-ESO Catalogue of Galactic Planetary 
Nebulae'' (Acker et al.\ 1992) at ftp://cdsarc.u-strasbg.fr/cats/V/84, 
and correlated the PN coordinates with the pointings of all archival 
ROSAT observations. 
In order to assure the highest quality for the data, we only selected  
those ROSAT observations with exposure times longer than 1,000 s. 
For the PSPC observations, we further restricted the selection to those in 
which the angular distances of the PNs to the field centers were less 
than {20\arcmin}. 
Eighty PNs\footnote{
M\,1-67 is in the Acker et al.'s (1992) list of PNs and has ROSAT 
HRI observations, but we do not include it because it has been 
shown to be a ring nebula around a Pop.~I Wolf-Rayet star (Crawford \& 
Barlow 1991). 
Similarly, we exclude EGB\,4 because it is likely produced 
by the strong wind from BZ\,Cam (Hollis et al.\ 1992), a white-dwarf star 
in a cataclysmic binary system (Kraft, Krzeminski, \& Mumford 1969). \\}
have ROSAT observations that meet these criteria. 
Of these, 55 have only PSPC observations, 8 have only HRI observations, and 
17 have both PSPC and HRI observations.

Table~1 lists the 80 PNs and their ROSAT observations. 
Columns~1 and 2 give the galactic coordinates and names of PNs. 
Column~3 gives the ROSAT observation sequence number: ``rp'' stands for 
PSPC observations without any filter, ``rf'' for PSPC observations with 
the boron filter, and ``rh'' for HRI observations. 
Columns~4 and 5 give the exposure time and the offset of the PN position 
from the pointing center. 
When two or more observations with the same instrument are available for a 
nebula, these observations have been aligned and merged into one file to 
increase the exposure time and to improve the signal-to-noise ratio. 
The merged files are used in our analysis.

Of these 80 nebulae, some are projected within or near a bright X-ray 
source and the identification of the PN emission is uncertain. 
Such is the case for: Ps\,1, the PN in the globular cluster M\,15; H\,4-1, 
in the direction of the Virgo Cluster; H\,2-12, Pe\,2-8, and Wray\,16-20, 
located in the direction of supernova remnants; and A\,12, H\,1-43, 
He\,2-51, and M\,1-30, too close to bright X-ray point sources. 
Therefore, we have excluded these objects from our final sample.

The final sample consists of 63 objects with PSPC observations (13 
have additional HRI observations), and 8 objects with only HRI 
observations. 
Given the limited spectral information and sensitivity of the HRI 
observations, we will focus the statistical study on the sample of 63 
PNs with PSPC observations. 
Hereafter, we will refer to this as the ROSAT PN sample.

\subsection{X-ray Emission from Planetary Nebulae}

To search for X-ray emission from PNs in the ROSAT sample, we have 
compared the PSPC X-ray images with the optical images extracted from 
the Digitized Sky Survey\footnote{
The Digitized Sky Survey is based on photographic data obtained using the 
UK Schmidt Telescope and the Oschin Schmidt Telescope on Palomar Mountain. 
The UK Schmidt was operate by the Royal Observatory of Edinburgh, with 
funding from the UK Science and Engineering Research Council, until 1988 
June, and thereafter by the Anglo-Australian Observatory. 
The Palomar Observatory Sky Survey was funded by the National Geographic 
Society. 
The Oschin Schmidt Telescope is operated by the California Institute of 
Technology and Palomar Observatory. 
The plates were processed into the present compressed digital form with 
the permission of these institutes. 
The Digitized Sky Survey was produced at the Space Telescope Science 
Institute under US government grant NAGW-2166.} (DSS). 
For each PN, we plot the X-ray contours over the optical image to 
determine whether X-ray emission is detected within the optical boundary 
of the nebula. 
We then define a source region that encompasses the entire nebula, use a 
surrounding annulus as a background region, prorate the background and 
compute the background-subtracted PSPC counts within the source region.  
For unresolved PNs, we use source regions that match the PSPC's
point spread function.

X-ray emission was detected in 13 PNs, of which NGC\,7009 is the only 
detection not previously reported in the literature. 
Their identifications and X-ray counts are listed in Table~2. 
Columns 1 and 2 give their Galactic coordinates and names, column 
3 the net PSPC counts detected within the nebulae, and column 4 the PSPC 
count rates. 
In order to facilitate comparisons with the non-detections, we have also 
listed the 3$\sigma$ values of the count rates in column 5.  
The X-ray spectral type (described below in this section) is given in 
column 6. 
Additional notes for some individual nebulae are given in Appendix~A.

Table 2 contains all PNs confirmed to emit X-rays, with the exception 
of NGC\,4361, which does not have pointed ROSAT observations.  
We note that X-ray emission is also tentatively detected in three 
additional PNs at a 2$\sigma$ level. 
These three PNs are listed in Table 3, in which information similar to that 
in Table 2 is given.  
Follow-up observations are needed to confirm or reject the X-ray emission 
from these three PNs. 
For the remaining 47 PNs that are not detected by ROSAT observations, we 
have determined their 3$\sigma$ detection limits and listed them in Table 4.

Figure~1 shows the distributions of exposure time and the 
3$\sigma$ detection limit of PSPC observations of the ROSAT PN sample. 
About 60\% of the detections were made in observations with $\le$10 ks
exposure time.  Many nondetections have very stringent 3$\sigma$ upper
limits.  Thus, it may be concluded that the data quality is
high and that the nondetections are caused by low X-ray fluxes from 
these PNs instead of short exposure times.

To illustrate the distribution of X-ray emission within the PNs,
Figure~2 displays X-ray images extracted from the PSPC 
observations alongside the DSS images extracted over the same
field of view and overlaid by the X-ray contours.  Figure~2
also shows two examples of PN misidentifications (A\,12 and A\,33).
It can be seen
that all 13 detected PNs have X-ray emission concentrated at the 
center of the nebula. None of the 13 X-ray sources in PNs are
``clearly" resolved. Only after painstakingly analyses can one
find that A\,30 and BD+30$^\circ$3639 are marginally resolved by
the HRI (Chu, Chang, \& Conway 1997; Leahy et al.\ 1998) and 
NGC\,6543 marginally resolved by the PSPC (Kreysing et al.\ 1992).
(By ``marginally resolved" we mean that the FWHM of the surface brightness 
profile is not larger than 1.5 times the instrumental FWHM.)

We have extracted X-ray spectra of these 13 PNs from their PSPC 
observations. 
The corresponding PSPC count rate plots are displayed in
Figure~3. 
These spectra can be roughly classified into three types, as noted by 
Conway \& Chu (1997): \begin{itemize}
\item 
Type 1 spectra have PSPC counts peaking near 0.1--0.2 keV and vanishing 
above 0.5 keV; this group includes NGC\,246, NGC\,1360, NGC\,3587, NGC\,6853, 
A\,30, and K\,1-16.
\item 
Type 2 spectra have PSPC counts peaking at $>$0.5 keV; this group includes 
NGC\,6543 and BD+30$^\circ$3639.  
\item 
Type 3 spectra have PSPC counts peaking strongly near 0.1--0.2 keV and 
weakly at 0.5--1.0 keV; this group includes NGC\,7293 and LoTr\,5.  
\end{itemize}
The spectra of NGC\,7009, A\,36 and K\,1-27, shown in Figure~4,
are too noisy to be classified definitively.  
As their detected photons have energies mostly above 0.5 keV and do not peak 
at 0.1--0.2 keV, it is most likely that these three spectra are of Type 2.

The three types of spectra can be fitted by different models.
Type 1 spectra can be fitted by blackbody models at temperatures of 
100,000 to 200,000 K or atmospheric models of similar stellar effective
temperatures (Chu et al.\ 1993; Hoare et al.\ 1995).
Type 2 spectra can be fitted by thin plasma emission models 
(e.g., Raymond \& Smith 1977) for a plasma temperature of a few 
10$^6$ K (Arnaud et al.\ 1996). 
Type 3 spectra can be fitted by the combination of a blackbody model and 
a thin plasma emission model (Leahy et al.\ 1994).

\section{Discussion}

\subsection{Detectability of X-ray Emission from Planetary Nebulae}

In the ROSAT PN sample, $\sim20\%$ of the PNs are detected; their 
PSPC count rates range from 0.002 to 0.4 counts~s$^{-1}$. 
The detection rate rises to $\sim25\%$, if the tentative detections 
for NGC\,2371-72, NGC\,2392, and NGC\,6572 are included. 
A similar detection rate, 25\%, can be derived from the Einstein and 
EXOSAT observations of PNs (Tarafdar \& Apparao 1988; Apparao \& 
Tarafdar 1989) if misidentifications are disregarded.

Note that the aforementioned detection rate should not be interpreted 
as the fraction of PNs that emit X-rays, because the detectability depends 
on different factors. 
These include not only the X-ray luminosity and distance of the PN, 
but also a complex interplay among the X-ray spectrum of the PN, the 
intervening absorption, and the energy-dependent response of the detector. 
In the ROSAT energy band, the interstellar absorption cross section is 
large, in particular at the low energy end.  Because the absorption is 
energy dependent, its effects on X-ray sources with different intrinsic
spectra can be drastically different.  To illustrate this, we have 
calculated the ratio of the absorbed and unabsorbed flux in the ROSAT 
PSPC band as a function of absorption column density, $N_{\rm H}$, for
two models of 
X-ray emission: a blackbody model for stellar photospheric emission,
and the Raymond \& Smith (1977) model for thin plasma emission. 
These two models are used because the blackbody model simulates a Type 1 
spectrum and the thin plasma emission model simulates a Type 2 spectrum.  
Morrison \& McCammon's (1983) effective cross section per hydrogen atom 
is used for the absorption, and solar abundances are used for the thin 
plasma emission.

Figure~5 shows the absorbed to unabsorbed flux 
ratios for blackbody emission at three temperatures appropriate for hot 
central stars of PNs (left panel) and for thin plasma emission at three 
plasma temperatures within $1-4\times10^{6}$ K (right panel).
The X-ray emission for a blackbody at 
$1-2\times10^{5}$ K is very soft and the interstellar absorption cross 
section is very high in this soft band; therefore, the blackbody emission 
is much more strongly absorbed by even a moderate column of intervening 
material than emission from a hot thin plasma.

To examine the range of absorption column densities commonly seen for 
PNs, we convert the optically determined extinction to an H{\sc i} 
column density, $N_{\rm H{\sc i}}$, and use it to approximate the absorption
column density, $N_{\rm H}$.  It should be born in mind that the absorption
column density, $N_{\rm H}$, consists of not only the atomic gas but also 
ionized gas and molecular gas, and therefore $N_{\rm H{\sc i}}$ provides 
only a lower limit for $N_{\rm H}$.  
A large number of PNs have $c_{\rm H\beta}$, the logarithmic extinction 
at H$\beta$, available in the literature (e.g., Cahn, Kaler, \& 
Stanghellini 1992; Tylenda et al.\ 1992).  
The extinction at H$\beta$ can be related to the
visual extinction A$_{\rm v}$ and to the color excess $E(B-V)$ by
A$_{\rm v}$ = 3.1~$E(B-V)$ = 2.17~$c_{\rm H\beta}$.  Using Bohlin, Savage, 
\& Drake's (1978) gas to dust ratio,
$N_{\rm H{\sc i}} / E(B-V) = 5.8\times10^{21}$ cm$^{-2}$ mag$^{-1}$,
we derive $N_{\rm H{\sc i}}$ from $c_{\rm H\beta}$.

Typical $c_{\rm H\beta}$ values of Galactic PNs range from a few tenths 
to greater than one, corresponding to A$_{\rm v}$ of 1--2 mag or 
$N_{\rm H{\sc i}}$ = $2-4\times10^{21}$ cm$^{-2}$.  Figure~5
illustrates
that blackbody emission would have been attenuated by factors of 10$^5$
to 10$^{10}$, while emission from a thin plasma at $2\times10^6$ K is 
attenuated by factors of a few tens.  It is conceivable that the
detectability of stellar photospheric emission is more heavily dependent
on the intervening absorption and that most PNs' photospheric emission
are attenuated below detection limit.

\subsection{Absorption Effects in the Survey Results}

In order to examine the absorption effects in the X-ray observations 
of the ROSAT PN sample, we have listed the distance ($d$), visual 
extinction ($A_{\rm v}$), and the corresponding H{\sc i} column density
($N_{\rm H{\sc i}}$) in Table 5.  Histograms of the distribution of PNs with
respect to $A_{\rm v}$, distance, and Galactic latitude and longitude
are presented in Figure~6.

It is clear from Figure~6 that the intervening
absorption column severely limits the detection of X-ray emission from PNs:
none in the ROSAT PN sample are detected with $A_{\rm v}$ greater than
1 mag.  Indeed, most of the detections have $A_{\rm v} \le 0.4$ mag.
The only object detected with a high extinction, $A_{\rm v} = 1$ mag,
is BD$+$30$^{\circ}$3639, which has a Type 2 spectrum and is best fitted 
by a $3\times10^6$ K thin plasma emission model (Arnaud et al.\ 1996).   
BD$+$30$^{\circ}$3639 can be detected because its X-ray emission is
attenuated by less than a factor of 10, as shown in 
Figure~5 (right panel).

The distribution of distances for the ROSAT PN sample is shown in
the upper right panel of Figure~6.  No PNs further than
1.7 kpc are detected.  As extinction usually increases with distance,
the lack of detection at large distances is not surprising.

The distributions of the ROSAT PNs with respect to Galactic latitude and
Galactic longitude, shown in the right panels of Figure~6,
also reflect the effects of absorption.   As most of the PNs are in
the Galactic plane, the high-latitude nebulae are nearby and have smaller
extinction, while the low-latitude nebulae are in general more distant
and have larger extinction.  It is thus expected that high-latitude nebulae
are easier to detect.  The only low-latitude nebula detected is the Dumbbell
Nebula, which is located at a small distance, 262 pc.  The distribution of 
the ROSAT PN sample peaks at Galactic longitude = 0.  As most of these nebulae 
are in the Galactic center with several magnitudes of extinction, their
non-detection can be expected.

\subsection{Physical Properties of PNs and X-ray Emission}

In order to understand the detectability and emission mechanisms of
X-ray emission from PNs, we have included angular radius 
($\theta$) and electron density ($N_{\rm e}$) of the main
nebula\footnote{We use the radius of the main nebula, instead of the 
large radius of the outermost features, such as the large halos around
BD+30\arcdeg3639 and NGC\,6543, because all X-ray sources in PNs are located 
within the main nebula and the dense main nebula may absorb soft X-rays 
significantly}, 
effective temperature ($T_{\rm eff}$) and surface gravity ($g$) of the 
central star in Table 5.   Histograms of the distribution of PNs with 
respect to nebular density and radius, and stellar effective temperature
and surface gravity are presented in Figure~7.

The distributions of PNs detected in X-rays with respect to nebular
density and radius (the upper panels of Figure~7) show 
that the 10 PNs with Type 1 or Type 3 spectra are larger (radius 
$>$ 0.2 pc) and more tenuous ($N_{\rm e}$ $<$ 300 cm$^{-3}$).  
As a PN evolves, its radius increases and its density decreases.
Since soft X-rays are easily absorbed, evolved PNs are more transparent
than younger PNs to the soft X-rays observed in a Type 1 or a Type 3
spectrum because evolved PNs have smaller nebular densities.
The three PNs with Type 2 spectra are smaller and denser.   These three
PNs can be detected because their X-ray spectra are harder and their X-ray
emission is less attenuated.

The distributions of PNs with respect to the effective temperature and
surface gravity of the central star (the lower panels of 
Figure~7) show that PNs with Type 1 X-ray spectra have
higher stellar effective temperatures and surface gravity.  This is 
expected for photospheric X-ray emission from the hot central stars.
On the other hand, PNs with Type 2 spectra have lower stellar effective
temperatures and surface gravity, indicating that their emission mechanism
must be different from photospheric emission.

\subsection{Physical Nature of X-ray Emission from PNs}

Two types of X-ray emission are detected from PNs: soft X-ray emission
whose spectra can be fitted by blackbody models for temperatures of
$1-2\times10^5$ K, and harder X-ray emission whose spectra can be fitted
by thin plasma emission models of temperatures a few 10$^6$ K. 
The soft X-ray emission is extremely susceptible to intervening absorption,
only sources with absorption column densities $\ll$10$^{21}$ cm$^{-2}$
have been detected.  The harder X-ray emission is not as attenuated 
as the soft X-ray emission, thus can be detected even if the absorption
column density reaches $2\times10^{21}$ cm$^{-2}$.

The soft X-ray emission is most likely photospheric emission from the hot
central stars.   As a PN ages, the nebula expands
and becomes more tenuous, while the stellar $T_{\rm eff}$ increases 
and the heavy elements in the stellar atmosphere settle. 
Therefore, evolved PNs with hot central stars and optically thin nebular 
shells produce detectable X-ray fluxes.
The PNs showing soft X-ray emission all have optically determined stellar
properties supporting the photospheric emission explanation.

There are a number of PNs in the ROSAT sample that have stellar 
effective temperature greater than $1\times10^5$ K but are not detected.  
These non-detections are mostly caused by high absorption, with $N_{\rm H}$ 
a few times 10$^{21}$ cm$^{-2}$, e.g., A\,21, Hb5, K3-92, M3-28, and NGC\,6565.
Only two non-detections have low absorption column density: A\,31 and NGC\,1535.
The non-detection of A\,31 is probably due to the insufficient integration time 
of the observation, only 1,023 s.   NGC\,1535 has a dense nebular shell,
and its non-detection is probably caused by the absorption in the dense shell.

There might be an additional mechanism to produce soft X-rays from PNs, as 
illustrated by A\,30, which shows only soft X-ray emission below 0.5 keV (Chu 
\& Ho 1995), but diffuse X-ray emission around the bright central source is 
detected at a 2$\sigma$ level (Chu et al.\ 1997).  A\,30 is a 
large, evolved, ``born-again" PN with a very hot central star (Iben et al.\ 
1983); therefore, the point X-ray source at the center is likely 
photospheric emission from the central star.  The marginally detected 
diffuse emission has a rough spatial correspondence with the radial
filaments and bipolar outflows, so it may be produced by interactions 
between the fast stellar wind and the H-deficient central nebula.
The soft spectrum of the diffuse emission implies a very low plasma
temperature, $< 10^6$ K (Chu et al.\ 1997).

The emission mechanism of the hard X-ray emission from PNs is 
largely uncertain. 
In Kwok et al.'s (1978) wind-wind interaction model, the heat 
conduction between the hot, shocked fast wind and the cold nebular 
shell would raise the density in the hot gas and produce detectable
X-ray emission (Soker 1994).
Using this mechanism, Zhekov \& Perinotto (1998) have reproduced
the spectral properties of NGC\,6543 satisfactorily.
However, the spatial resolution of the available ROSAT PSPC 
observations is inadequate to confirm this mechanism morphologically.
Among the five PNs showing hard X-ray emission, BD+30\arcdeg3639, 
NGC\,6543, NGC\,7009, NGC\,7293, and LoTr\,5, only the first two
have been marginally resolved by the ROSAT HRI and PSPC, respectively.
The X-ray emission might originate within the star, e.g., in the corona 
(Fleming, Werner, \& Barstow 1993) or even in an X-ray binary. 
Future high-resolution observations by the Chandra Observatory would be
very useful in determining the emission mechanism of hard X-rays.

Of the five PNs with hard X-ray emission, two also have soft X-ray emission
(i.e., Type 3 spectra), LoTr\,5 and NGC\,7293.  These two nebulae, like the 
PNs with only soft X-ray emission, are large, tenuous, evolved PNs with hot 
central stars and low extinction.   On the other hand, the three nebulae with 
only hard X-ray emission (i.e., Type 2 spectra) are small, dense, and young, 
and their central stars are well below $1\times10^5$ K.   
X-ray luminosities of three PNs with hard X-ray emission have been reported: 
BD+30\arcdeg3639 (Type 2) has an X-ray luminosity of $2-3\times10^{32}$ 
erg s$^{-1}$ in the 0.4--1.7 keV band (Arnaud et al.\ 1996), 
NGC\,6543 (Type 2) $3\times10^{31}$ erg s$^{-1}$ in the 0.1--2.4 keV band 
(Kreysing et al.\ 1992),
and NGC\,7293's (Type 3) hard component $2.6\times10^{29}$ erg s$^{-1}$ 
in the 0.1--2.4 keV band (Leahy et al.\ 1994).
It is not clear whether the hard X-ray emission mechanism is the same for 
these two groups of PNs with contrasting physical properties.
It is nevertheless clear that Type 2 spectra could not have resulted 
from heavily absorbed Type 3 spectra because the central stars of 
PNs with Type 2 spectra are too cool to produce soft X-ray emission.

As the hard X-rays are not as severely attenuated as the soft X-rays, the
small number of hard X-ray sources detected in PNs indicates that hard
X-ray emission from PNs is not universal as is the case of soft X-ray
emission from a hot photosphere.  To understand the nature of hard X-ray
emission from PNs, sensitive searches for more sources are needed, and
follow-up observations with high spatial and spectral resolution are
particularly important.

\section{Summary and Conclusions}

We have used the ROSAT PSPC and HRI archival observations to study X-ray 
emission from PNs.  63 PNs have useful ROSAT PSPC observations.  These
observations are used to analyze the spatial and spectral properties of 
the X-ray emission for positive detections and to determine the 3$\sigma$ 
upper limits for non-detections.  We have examined the effects of 
absorption and use them to explain the distribution of PNs with detectable
X-ray emission with respect to distance, extinction, Galactic latitude
and longitude, effective temperature and surface gravity of the central
star, and size and density of the nebular shell.
The main results are summarized below. 

\begin{enumerate}
\item $20-25\%$ of the PNs with useful ROSAT PSPC observations show 
      X-ray emission.   These include 13 definite detections and 3 
      possible detections (at a 2$\sigma$ level).
\item Three types of spectra are seen in PNs: Type 1 consists of only
      soft X-ray emission ($<$0.5 keV), peaks at 0.1--0.2 keV, and
      can be fitted by blackbody models at temperatures $1-2\times10^5$ K;
      Type 2 consists of harder X-ray emission, peaks at $>$0.5 keV, and
      can be fitted by thin plasma emission models at temperatures of
      a few 10$^6$ K; Type 3 is a composite of a bright Type 1 component 
      and a fainter Type 2 component.
\item The unresolved soft X-ray emission (Type 1 spectra) is most likely 
      photospheric emission from hot central stars.  As soft X-ray 
      emission can be severely attenuated by intervening
      absorption, only large, evolved PNs with hot central
      stars and small absorption column density,  
      $\ll1\times10^{21}$ cm$^{-2}$, are detected.
\item Marginally resolved soft X-ray emission is detected in the
      evolved, H-deficient PN A\,30.  Its central bright point X-ray
      source is probably photospheric emission, and the surrounding
      diffuse emission (at a 2$\sigma$ level) may result from the
      interaction between the fast stellar wind and the surrounding medium.
\item The origin of the hard X-ray emission (Type 2 spectra and the hard 
      component of Type 3 spectra) is uncertain.  It could originate in
      a stellar corona, X-ray binary, or shocked stellar wind.
      As hard X-ray emission is not as easily attenuated as soft X-ray
      emission, the small number of hard X-ray emitting PNs 
      implies a real paucity of hard X-ray sources among PNs.
\item PNs with Type 2 spectra and PNs with Type 3 spectra exhibit 
      different stellar and nebular properties.  The former are
      associated with small, dense, young PNs with relatively cool ($\ll10^5$ K)
      central stars, while the latter are associated with large, tenuous,
      evolved
      PNs with hot ($\le10^5$ K) central stars.  These two groups of PNs
      may have different emission mechanisms for their hard X-rays.  
\item Among PNs showing hard X-ray emission (Type 2 or Type 3), only two are 
      marginally resolved, BD+30\arcdeg3639 and NGC\,6543, both with 
      Type 2 spectra.
      The X-ray luminosities of these PNs are 10$^2-10^3$ times higher than
      that of the hard component of NGC\,7293, whose spectrum is of Type 3.
      Observations with high spatial resolution are necessary to 
      determine the origin of their hard X-ray emission.
\end{enumerate}	 

\acknowledgments
 
This work is supported by the NASA grant NAG 5-8103 (ADP).  MAG is 
supported partially by the Direcci\'on General de Ense\~nanza 
Superior e Investigaci\'on Cient\'{\i}fica of the Spanish Ministry 
of Education and Culture. 
This research has made use of data obtained through the High Energy 
Astrophysics Science Archive Research Center Online Service, provided 
by the NASA/Goddard Space Flight Center.

\appendix

\section{Notes on Individual Objects}

\noindent{\bf PN G 036.1--57.1 = NGC\,7293 = Helix Nebula}

Despite the large angular size ($\sim960\arcmin$) of the nebula,
its X-ray emission is unresolved and coincident with the position
of the central star.  Its X-ray spectrum, Type 3, cannot be fitted 
by one single component; two components are needed.  The soft 
component is most likely the photospheric emission from the hot 
central star, but the origin of the hard component is less certain
(Leahy et al.\ 1994). 

\noindent{\bf PN G 037.7--34.5 = NGC\,7009 = Saturn Nebula}

The Saturn Nebula has an elliptical ($25\arcsec\times10\arcsec$) main 
nebula with two ansae extending out to {25\arcsec} from the central star. 
The total number of PSPC counts detected is too low to determine
reliably whether the X-ray emission is from a point source or is extended.
While the PSPC spectrum is too noisy for spectral fits, it shows that most
of the detected X-ray photons have energies greater than 0.5 keV, 
indicating that the X-ray emission probably originates from a hot plasma, 
instead of the stellar photosphere.  Future observations with higher
spatial and spectral resolution are needed to determine the real nature of
the Saturn Nebula's X-ray emission.

\noindent{\bf PN G 060.8--03.6 = NGC\,6853 = Dumbbell Nebula}

The Dumbbell Nebula is the brightest X-ray source among PNs known to 
emit X-rays.
Its X-ray emission appears extended in PSPC images extracted from energies
below 0.2 keV (Kreysing et al.\ 1992) owing to an electronic ghost image
problem (Chu et al.\ 1993).  The PSPC images extracted at energies greater
than 0.2 keV and the HRI image both show only a point source at the position 
of the central star.   The PSPC spectrum peaks at the lowest energy bins.
These spatial and spectral properties indicate that the X-ray emission from
the Dumbbell Nebula originates in the central star's photosphere.

\noindent{\bf PN G 064.7+05.0 = BD+30\arcdeg3639}

BD+30\arcdeg3639 is very likely a young PN because of its small 
($\sim4\arcsec$ = 0.04 pc), dense nebular shell and the low temperature 
of its central star (Harrington et al.\ 1997).  Its X-ray spectrum can 
be fitted by a thin plasma emission model for a temperature of 
$3\times10^6$ K (Kreysing et al.\ 1992; Arnaud et al.\ 1996).  The observed
X-rays cannot be explained by stellar photospheric emission.  Leahy et 
al.\ (1998) analyzed an HRI image of BD+30\arcdeg3639 and reported that the 
X-ray emission is marginally extended for a resolution of $\sim5\arcsec$.  
It is possible that the X-ray emission of BD+30\arcdeg3639 is from the 
hot plasma in the central cavity of the PN shell, but high-resolution
observations in the future are needed to confirm this result.

\noindent{\bf PN G 094.0+27.4 = K\,1-16}

K\,1-16 is an extended ($90\arcsec\times70\arcsec$) low surface brightness 
nebula (Manchado et al.\ 1996) with a He-rich PG1159 type central star.
Its X-ray spectral distribution is compatible with the stellar 
photospheric emission of its hot central star (Hoare et al.\ 1995).

K\,1-16 is projected within a cluster of galaxies and $\sim2\arcmin$
from the QSO E1821+643 (Saxton et al.\ 1997).   As the X-ray emission 
from the PN is superposed on the extended emission from the cluster of 
galaxies, the extraction of X-ray emission from the PN is somewhat
difficult.  For background subtraction, we have chosen background
regions with similar surface brightness around the QSO.  The resultant
X-ray spectrum of K\,1-16 (Figure~3) does not show any 
significant emission at 1.0 keV, where the QSO spectrum peaks. 
Therefore, the background subtraction is probably reasonable and the
QSO contamination of the PN spectrum at the soft 0.1--0.3 keV band is 
probably negligible. 

\noindent{\bf PN G 096.4+29.9 = NGC\,6543 = Cat's Eye}

ROSAT PSPC observations of this PN show extended X-ray emission, and the
PSPC spectrum is best fitted by thin plasma emission models for 
temperatures of a few 10$^6$ K (Kreysing et al.\ 1992).  This nebula
is thus the best candidate for follow-up study of diffuse X-ray emission
from PNs.  The optical size of the bright central nebula 
($\sim26\arcsec\times18\arcsec$) matches roughly the PSPC resolution. 
High-resolution observations with Chandra should reveal clearly the
distribution of X-ray emission and its relationship with the optical
morphological features, allowing us to determine the nature of the
X-ray emission.

\noindent{\bf PN G 118.8--74.7 = NGC\,246}

NGC\,246 is a very extended ($\sim5\arcmin\times6\arcmin$) low surface 
brightness nebula.  Its central star is also a member of the He-rich PG1159 
class of pre-white dwarfs, similar to the central star of K\,1-16.  The
central star of NGC\,246 is a binary (Cudworth 1973) with a companion 
widely separated by 1900 AU (Howard \& Ciardullo 1999).  The X-ray emission 
from NGC\,246 is consistent with that expected from the photosphere of the
hot central star.   It is unnecessary to assume an X-ray binary to explain 
the X-ray emission.

\noindent {\bf PN G 143.6+23.8 = EGB 4}

As already noted in $\S$2, the PN classification for EGB\,4 is uncertain. 
Bright X-ray emission is detected from its central star BZ Cam, a white
dwarf in a cataclysmic binary system.  The spectral distribution of the 
X-ray emission is consistent with such a binary system (van Teeseling \& 
Verbunt 1994).

\noindent{\bf PN G 148.4+57.0 = NGC\,3587 = Owl Nebula}

The PSPC X-ray spectrum of the Owl Nebula was reported by Leahy et al.\ (1996);
however, the double-hump spectral shape was caused by the inclusion of a 
background source.  ROSAT HRI observations of the Owl Nebula clearly show
two point sources projected within the nebula, and both sources have a stellar
counterpart.   One source is coincident with the position of the central star
and its PSPC spectrum shows only photons with energies below 0.5 keV, 
consistent with the photospheric emission expected from the hot central star.
The other source is probably associated with a background star; its PSPC
spectrum is similar to those of stellar coronal emission.  A detailed 
description of the X-ray sources has been reported by Chu et al.\ (1998).

\noindent{\bf PN G 198.6--06.3 = A\,12}

The X-ray emission from A\,12 reported by Kreysing et al.\ (1992) was
misidentified.   As shown in Figure~2 and pointed out by
Hoare et al.\ (1995), the X-ray source is coincident in position with 
the bright binary star $\mu$ Ori located $\sim1\arcmin$ from A\,12 (Hoare 
et al.\ 1995).  No X-ray emission is evident at the position of A\,12.

\noindent{\bf PN G 206.4--40.5 = NGC\,1535}

X-ray emission from NGC\,1535 has been reported by Apparao \& Tarafdar 
(1989) using EXOSAT observations and by Leahy et al.\ (1996) using
ROSAT PSPC observations.   A comparison
between optical and PSPC X-ray images reveals that the X-ray source is
outside the optical boundary of NGC\,1535, and appears to be associated 
with a field star.  The previously reported X-ray emission from NGC\,1535 
is therefore mis-identified; no X-ray emission is detected within the 
optical boundary of NGC\,1535 (Chu et al.\ 1998).

\noindent{\bf PN G 208.5+33.2 = A\,30}

A\,30 contains a H-deficient central nebula and a large halo.
The X-ray emission was serendipitously discovered by Chu \& Ho (1995)
using a ROSAT PSPC observation centered on a cluster of galaxies.
Subsequent HRI observations confirm the detection and reveal a marginally
extended X-ray source at the central H-deficient nebula (Chu et al.\ 1997).
Unfortunately, the X-ray emission from A\,30 is too faint and too soft
for Chandra, and is too small for XMM.

\noindent{\bf PN G 220.3--53.9 = NGC\,1360}

NGC\,1360 is the first PN reported to emit X-rays (de Korte et al.\
1985).  Its optical nebula is extended ($\sim5.5\arcmin\times3\arcmin$) 
and has a low surface brightness.  The unresolved X-ray emission from 
NGC\,1360 is coincident in position with the central star, and the X-ray
spectrum can be modeled as photospheric emission from a hot star
(Hoare et al.\ 1995).

\noindent{\bf PN G 238.0+34.8 = A\,33}

The X-ray emission from A\,33 was reported by Tarafdar \& Apparao (1988) 
using Einstein observations.  As in the cases of A\,12 and NGC\,1535,
a comparison between optical and PSPC X-ray images (see Figure 2) shows 
that the X-ray source is outside the optical nebula of A\,33.  
The previously reported X-ray emission from A\,33 is therefore spurious 
(Conway \& Chu 1997).

\noindent{\bf PN G 286.8--29.5 = K\,1-27}

K\,1-27 is an elliptical ($65\arcsec\times50\arcsec$) PN with a low
surface brightness.  The total number of PSPC counts detected is too
low to allow detailed spatial or spectral analysis.  The PSPC spectrum 
does not peak in the lowest energy channels as have other spectra
attributed to photospheric emission from hot stars (e.g., NGC\,246 and 
NGC\,1360).   Deeper observations with higher spatial resolution are 
needed to determine the spatial distribution and spectral properties of 
the X-ray emission from K\,1-27.

\noindent{\bf PN G 318.4+41.4 = A\,36}

A\,36 is an extended ($\sim8\arcmin\times5\arcmin$) nebula with a low 
surface brightness.  Leahy et al.\ (1996) reported the PSPC X-ray 
spectrum of A\,36; however, their extracted spectrum included the
contribution of a background source projected near the northern boundary 
of the nebula (see Figure~2).  Using a small aperture that 
includes only the X-ray source at the central star of A\,36, we have
extracted a PSPC spectrum.  This spectrum shows only photons with 
energies below 1 keV, but does not peak in the lowest energy channels as 
the spectra from objects interpreted as photospheric emission from hot 
stars.  Deeper observations are needed to determine the spatial and 
spectral properties of the X-ray emission from A\,36.

\noindent{\bf PN G 339.9+88.4 = LoTr\,5}

LoTr~5 is an extended ($\sim17\arcmin$), nearby PN of extremely low 
surface brightness. 
Its central star is known to be a binary (perhaps even triple) system 
(Jasniewicz et al.\ 1996, and references therein). 
The X-ray spectrum of LoTr\,5 is similar to that of the Helix Nebula,
consisting of a strong soft component and a weak hearder component.

\clearpage

\onecolumn

\begin{deluxetable}{lllcc}
\tablenum{1}
\tablewidth{25pc}
\tablecaption{ROSAT PSPC and HRI point observations of PNe}
\tablehead{
\multicolumn{1}{l}{PN G} &
\multicolumn{1}{l}{PN Name} &
\multicolumn{1}{l}{ROSAT Obs.} &
\multicolumn{1}{c}{Exp. Time} &
\multicolumn{1}{c}{Offset} \\ 
\multicolumn{1}{l}{} &
\multicolumn{1}{l}{} &
\multicolumn{1}{l}{} &
\multicolumn{1}{c}{[s]} &
\multicolumn{1}{c}{[$\prime$]}
}

\startdata
000.1$-$01.1 & M 3-43    & rp400201n00 & ~~2202 &  11 \nl
000.1$+$02.6 & Al 2-J    & rp900204n00 & ~~3662 &  ~6 \nl
           &             & rh900394n00 & ~~4901 &  ~5 \nl
000.5$-$01.6 & Al 2-Q    & rp400209n00 & ~~2349 &  19 \nl
000.6$-$01.3 & Bl 3-15   & rp400209n00 & ~~2349 &  ~4 \nl
000.7$-$03.7 & M 3-22    & rp600418n00 &  29578 &  20 \nl
000.8$-$01.5 & Bl O      & rp400209n00 & ~~2349 &  19 \nl
001.2$-$03.9 & ShWi 2-5  & rp600418n00 &  29578 &  10 \nl
002.2$+$00.5 & Te 2337   & rp900199n00 & ~~4716 &  14 \nl
           &             & rp900199a01 & ~~1784 &  14 \nl
002.6$-$03.4 & M 1-37    & rp201093n00 &  10033 &  ~8 \nl
003.3$-$07.5 & KFL 19    & rp400054m01 & ~~5296 &  ~7 \nl
003.5$-$04.6 & NGC 6565  & rp300188n00 & ~~9980 &  20 \nl
003.8$-$04.3 & H 1-59    & rp300188n00 & ~~9980 &  15 \nl
003.9$+$01.6 & Te 2111   & rh300060n00 &  22543 &  ~7 \nl
004.5$+$06.8 & H 2-12    & rp500050n00 & ~~1784 &  ~1 \nl
           &             & rh500099n00 &  36662 &  ~1 \nl
007.8$-$03.7 & M 2-34    & rp500009n00 & ~~3992 &  ~5 \nl
009.8$-$07.5 & GJJC 1    & rp300058n00 & ~~8295 &  ~0 \nl
           &             & rh300184a01 &  31709 &  ~0 \nl
019.9$+$00.9 & M 3-53    & rp400326n00 &  11796 &  10 \nl
021.7$-$00.6 & M 3-55    & rp500287n00 & ~~9100 &  ~5 \nl
           &             & rp500287a01 & ~~5996 &  ~5 \nl
021.8$-$00.4 & M 3-28    & rp500287n00 & ~~9100 &  ~1 \nl
031.9$-$00.3 & WeSb 4    & rp900402n00 &  23825 &  19 \nl
034.6$+$11.8 & NGC 6572  & rp200673n00 & ~~4442 &  ~0 \nl
           &             & rh200145n00 & ~~3463 &  ~0 \nl
036.1$-$57.1 & NGC 7293  & rp900187n00 & ~~4914 &  ~0 \nl
037.7$-$34.5 & NGC 7009  & rp200672n00 & ~~4367 &  ~0 \nl
           &             & rh200141n00 & ~~3772 &  ~0 \nl
039.5$-$02.7 & M 2-47    & rp500058a02 &  20575 &  17 \nl
           &             & rp500058a01 &  18484 &  17 \nl
           &             & rp500058a00 & ~~5010 &  17 \nl
040.4$-$03.1 & K 3-30    & rp200126n00 & ~~7194 &  ~8 \nl
043.1$+$37.7 & NGC 6210  & rh200144n00 & ~~5611 &  ~0 \nl
049.3$+$88.1 & H 4-1     & rp800006n00 &  21545 &  10 \nl
053.3$+$03.0 & A 59      & rh201632n00 & ~~1752 &  ~3 \nl
053.8$-$03.0 & A 63      & rp201100n00 & ~~6726 &  20 \nl
060.8$-$03.6 & NGC 6853  & rp900016n00 & ~~5905 &  ~0 \nl
           &             & rp200568n00 & ~~5242 &  ~0 \nl
           &             & rp200568a01 & ~~5155 &  ~0 \nl
           &             & rp200568a02 & ~~3621 &  ~0 \nl
           &             & rh200569a01 & ~~3573 &  ~0 \nl
061.4$-$09.5 & NGC 6905  & rh201101n00 &  17578 &  ~0 \nl
064.7$+$05.0 & BD+30 3639 & rp201096n00 & ~~5834 &  ~0 \nl
             &            & rp500327n00 & ~~3686 &  17 \nl
             &            & rh900709n00 & ~~8740 &  ~0 \nl
065.0$-$27.3 & Ps 1      & rp400081n00 & ~~8780 &  ~0 \nl
           &             & rh400611n00 &  52468 &  ~0 \nl
068.7$+$01.9 & K 4-41    & rp201102n00 & ~~7941 &  ~1 \nl
068.7$+$03.0 & PC 23     & rp400048a00 & ~~5065 &  15 \nl
076.3$+$01.1 & A 69      & rp200062n00 &  13811 &  19 \nl
           &             & rp200063n00 & ~~2055 &  19 \nl
           &             & rp200058n00 & ~~1913 &  19 \nl
           &             & rp200057n00 & ~~1826 &  19 \nl
077.5$+$03.7 & KjPn 1    & rp500208n00 & ~~2302 &  13 \nl
077.7$+$03.1 & KjPn 2    & rp500207n00 & ~~9310 &  19 \nl
088.7$+$04.6 & K 3-78    & rp500220n00 & ~~4895 &  18 \nl
094.0$+$27.4 & K 1-16    & rp700413n00 & ~~1128 &  ~0 \nl
           &             & rp700948n00 & ~~1980 &  ~0 \nl
           &             & rp700949n00 & ~~1858 &  ~0 \nl
           &             & rh800754n00 &  29427 &  ~1 \nl
096.4$+$29.9 & NGC 6543  & rp000026n00 &  44989 &  ~9 \nl
           &             & rp170075n00 &  41204 &  15 \nl
           &             & rh120030n00 &  47482 &  ~9 \nl
104.2$-$29.6 & Jn 1      & rp200122n00 & ~~1140 &  ~0 \nl
           &             & rf200122n00 & ~~4192 &  ~0 \nl
110.6$-$12.9 & K 1-20    & rp900343n00 & ~~8838 &  16 \nl
118.8$-$74.7 & NGC 246   & rf200842n00 & ~~7471 &  ~0 \nl
           &             & rf200842a01 & ~~3962 &  ~0 \nl
130.4$+$03.1 & K 3-92    & rp180068n00 & ~~5826 &  14 \nl
148.4$+$57.0 & NGC 3587  & rp900186n00 & ~~1783 &  ~0 \nl
           &             & rh900706n00 &  35224 &  ~0 \nl
166.4$-$06.5 & CRL 618   & rh201614n00 &  18697 &  ~0 \nl
183.8$+$05.5 & WeSb 2    & rp300191n00 &  17255 &  14 \nl
189.1$+$19.8 & NGC 2371-72 & rp200450n00 &  11758 &  ~0 \nl
197.8$+$17.3 & NGC 2392  & rp500112n00 & ~~3795 &  19 \nl
198.6$-$06.3 & A 12      & rp201097n00 &  10287 &  ~0 \nl
           &             & rh201866n00 & ~~4923 &  ~0 \nl
205.1$+$14.2 & A 21      & rp200121a01 & ~~1628 &  ~0 \nl
           &             & rp200121a00 & ~~1192 &  ~0 \nl
206.4$-$40.5 & NGC 1535  & rp900242n00 &  11620 &  ~0 \nl
208.5$+$33.2 & A 30      & rp800370n00 &  21773 &  11 \nl
           &             & rh900643n00 &  55795 &  ~0 \nl
215.2$-$24.2 & IC 418    & rp200676n00 & ~~5085 &  ~0 \nl
           &             & rh200671n00 & ~~6040 &  ~0 \nl
219.1$+$32.2 & A 31      & rp200120n00 & ~~1023 &  ~0 \nl
220.3$-$53.9 & NGC 1360  & rf200843n00 & ~~8722 &  ~0 \nl
233.5$-$16.3 & A 15      & rp201071n00 & ~~4439 &  ~3 \nl
238.0$+$34.8 & A 33      & rp200119a01 & ~~1526 &  ~1 \nl
           &             & rp200119n00 & ~~1205 &  ~1 \nl
260.7$-$03.3 & Wray 16-20  & rp500055a01 & ~~1725 &  10 \nl
           &             & rp500055n00 & ~~1215 &  10 \nl
           &             & rh500164n00 & ~~3822 &  ~0 \nl
285.4$+$01.5 & Pe 1-1    & rp400394n00 & ~~1196 &  ~8 \nl
286.8$-$29.5 & K 1-27    & rp201245n00 &  18447 &  ~0 \nl
288.8$-$05.2 & He 2-51   & rp200868n00 & ~~2676 &  15 \nl
           &             & rp200863n00 & ~~2264 &  12 \nl
296.5$-$06.9 & He 2-71     & rh180140n00 & ~~1174 &  11 \nl
           &             & rh180141n00 & ~~2155 &  15 \nl
318.4$+$41.4 & A 36      & rp900244n00 & ~~9992 &  ~0 \nl
322.4$-$00.1 & Pe 2-8    & rp500160n00 & ~~4770 &  ~5 \nl
326.1$-$01.9 & VBe 3     & rp500010n00 & ~~3904 &  11 \nl
339.9$+$88.4 & LoTr 5    & rp201514n00 &  18788 &  ~0 \nl
346.3$-$06.8 & Fg 2      & rp400077n00 &  20937 &  17 \nl
347.7$+$02.0 & Vd 1-8    & rh400706n00 & ~~2467 &  ~8 \nl
352.9$-$07.5 & Fg 3      & rp300042n00 & ~~5605 &  11 \nl
355.9$-$04.2 & M 1-30    & rp201383n00 &  28760 &  14 \nl
           &             & rp200983a01 &  26361 &  15 \nl
           &             & rp200983n00 &  22287 &  15 \nl
357.1$-$04.7 & H 1-43    & rp400269n00 &  21168 &  ~5 \nl
357.6$+$01.0 & TrBr 4    & rh300062n00 & ~~2638 &  ~9 \nl
358.8$-$00.0 & Te 2022   & rp400275n00 &  15517 &  17 \nl
           &             & rp400275a01 &  13229 &  17 \nl
359.2$+$01.2 & 19W32     & rp400186n00 & ~~2100 &  ~9 \nl
359.3$+$01.4 & Th 3-35   & rp400186n00 & ~~2100 &  ~5 \nl
           &             & rh400747n00 &  11922 &  14 \nl
359.3$-$00.9 & Hb 5      & rp400179n00 & ~~2082 &  ~6 \nl
           &             & rh400177n00 & ~~3675 &  ~1 \nl
359.8$+$02.4 & Th 3-33   & rp900204n00 & ~~3662 &  16 \nl
359.8$-$07.2 & M 2-32    & rp200916n00 & ~~4581 &  11 
\enddata
\end{deluxetable}

\clearpage

\begin{deluxetable}{llrlll}
\tablenum{2}
\tablewidth{0pc}
\tablecaption{PNe with detected x-ray emission in ROSAT PSPC observations}
\tablehead{
\multicolumn{1}{c}{PN G} &
\multicolumn{1}{l}{PN Name} & 
\multicolumn{1}{l}{Count number} & 
\multicolumn{1}{c}{X-ray flux} & 
\multicolumn{1}{c}{3-$\sigma$} & 
\multicolumn{1}{c}{Spectral Type} \\ 
\multicolumn{1}{l}{} &
\multicolumn{1}{l}{} &
\multicolumn{1}{c}{[counts]} &
\multicolumn{1}{c}{[count s$^{-1}$]} &
\multicolumn{1}{c}{[count s$^{-1}$]} & 
\multicolumn{1}{l}{} 
}

\startdata
036.1$-$57.1 & NGC 7293    &  300$\pm$20~~~~  & ~~~~0.06   & ~~~~0.012  & ~~~Typ
e 3 \nl
037.7$-$34.5 & NGC 7009    &   24$\pm$~~7~~~~ & ~~~~0.005  & ~~~~0.005  & ~~~Typ
e 2 ? \nl
060.8$-$03.6 & NGC 6853    & 7700$\pm$90~~~~  & ~~~~0.39   & ~~~~0.014  & ~~~Typ
e 1 \nl
064.7$+$05.0 & BD+30 3639  &  420$\pm$25~~~~  & ~~~~0.044  & ~~~~0.007  & ~~~Typ
e 2 \nl
094.0$+$27.4 & K 1-16      &  200$\pm$17~~~~  & ~~~~0.04   & ~~~~0.010  & ~~~Typ
e 1 \nl
096.4$+$29.9 & NGC 6543    &  560$\pm$30~~~~  & ~~~~0.0065 & ~~~~0.0011 & ~~~Typ
e 2 \nl
118.8$-$74.7 & NGC 246     & 1050$\pm$35~~~~  & ~~~~0.091  & ~~~~0.009  & ~~~Typ
e 1 \nl
148.4$+$57.0 & NGC 3587    &   14$\pm$~~5~~~~ & ~~~~0.008  & ~~~~0.009  & ~~~Typ
e 1 \nl
208.5$+$33.2 & A 30        &  109$\pm$16~~~~  & ~~~~0.005  & ~~~~0.0022 & ~~~Typ
e 1 \nl
220.3$-$53.9 & NGC 1360    &  111$\pm$16~~~~  & ~~~~0.013  & ~~~~0.006  & ~~~Typ
e 1 \nl
286.8$-$29.5 & K 1-27      &   34$\pm$10~~~~  & ~~~~0.0018 & ~~~~0.0016 & ~~~Typ
e 2 ? \nl
318.4$+$41.4 & A 36        &   31$\pm$~~8~~~~ & ~~~~0.003  & ~~~~0.0023 & ~~~Typ
e 2 ? \nl
339.9$+$88.4 & LoTr 5      &  420$\pm$25~~~~  & ~~~~0.022  & ~~~~0.004  & ~~~Typ
e 3 \nl
\enddata
\end{deluxetable}

\begin{deluxetable}{llrll}
\tablenum{3}
\tablewidth{28pc}
\tablecaption{PNe tentatively detected in ROSAT PSPC observations}
\tablehead{
\multicolumn{1}{c}{PN G} &
\multicolumn{1}{l}{PN Name} & 
\multicolumn{1}{l}{Count number} & 
\multicolumn{1}{c}{X-ray flux} & 
\multicolumn{1}{c}{3-$\sigma$} \\ 
\multicolumn{1}{l}{} &
\multicolumn{1}{l}{} &
\multicolumn{1}{c}{[counts]} &
\multicolumn{1}{c}{[count s$^{-1}$]} &
\multicolumn{1}{c}{[count s$^{-1}$]}
}

\startdata
034.6$+$11.8 & NGC 6572    &   8$\pm$~5~~~~ & ~~~~0.0018 & ~~~~0.003 \nl  
189.1$+$19.8 & NGC 2371-72 &  23$\pm$14~~~~ & ~~~~0.0020 & ~~~~0.004 \nl
197.8$+$17.3 & NGC 2392    &   8$\pm$~5~~~~ & ~~~~0.0021 & ~~~~0.004 \nl     
\enddata
\end{deluxetable}
 
\begin{deluxetable}{lll|lll}
\tablenum{4}
\tablewidth{36pc}
\tablecaption{PNe undetected in ROSAT PSPC observations}
\tablehead{
\multicolumn{1}{c}{PN G} &
\multicolumn{1}{l}{PN Name} &
\multicolumn{1}{c}{3-$\sigma$} & 
\multicolumn{1}{c}{PN G} &
\multicolumn{1}{l}{PN Name} &
\multicolumn{1}{c}{3-$\sigma$} \\ 
\multicolumn{1}{l}{} &
\multicolumn{1}{l}{} &
\multicolumn{1}{c}{[count s$^{-1}$]} &
\multicolumn{1}{l}{} &
\multicolumn{1}{l}{} &
\multicolumn{1}{c}{[count s$^{-1}$]}
}
 
\startdata
000.1$-$01.1 & M 3-43    & ~~~0.004   & 077.5$+$03.7 & KjPn 1      & ~~~0.0025
\nl
000.1$+$02.6 & Al 2-J    & ~~~0.003   & 077.7$+$03.1 & KjPn 2      & ~~~0.0012
\nl
000.5$-$01.6 & Al 2-Q    & ~~~0.003   & 088.7$+$04.6 & K 3-78      & ~~~0.0014
\nl
000.6$-$01.3 & Bl 3-15   & ~~~0.006   & 104.2$-$29.6 & Jn 1        & ~~~0.03  
\nl
000.7$-$03.7 & M 3-22    & ~~~0.0006  & 110.6$-$12.9 & K 1-20      & ~~~0.0015
\nl
000.8$-$01.5 & Bl O      & ~~~0.004   & 130.4$+$03.1 & K 3-92      & ~~~0.0012
\nl
001.2$-$03.9 & ShWi 2-5  & ~~~0.0009  & 183.8$+$05.5 & WeSb 2      & ~~~0.0024
\nl
002.2$+$00.5 & Te 2337   & ~~~0.0020  & 205.1$+$14.2 & A 21        & ~~~0.03  
\nl
002.6$-$03.4 & M 1-37    & ~~~0.0015  & 206.4$-$40.5 & NGC 1535    & ~~~0.0016
\nl
003.3$-$07.5 & KFL 19    & ~~~0.0022  & 215.2$-$24.2 & IC 418      & ~~~0.0022
\nl
003.5$-$04.6 & NGC 6565  & ~~~0.0013  & 219.1$+$32.2 & A 31        & ~~~0.04  
\nl
003.8$-$04.3 & H 1-59    & ~~~0.0017  & 233.5$-$16.3 & A 15        & ~~~0.0016
\nl
007.8$-$03.7 & M 2-34    & ~~~0.003   & 238.0$+$34.8 & A 33        & ~~~0.007 
\nl
009.8$-$07.5 & GJJC 1    & ~~~0.0020  & 285.4$+$01.5 & Pe 1-1      & ~~~0.007 
\nl
019.9$+$00.9 & M 3-53    & ~~~0.0012  & 326.1$-$01.9 & VBe 3       & ~~~0.003 
\nl
021.7$-$00.6 & M 3-55    & ~~~0.0008  & 346.3$-$06.8 & Fg 2        & ~~~0.0014
\nl
021.8$-$00.4 & M 3-28    & ~~~0.0014  & 352.9$-$07.5 & Fg 3        & ~~~0.0024
\nl
031.9$-$00.3 & WeSb 4    & ~~~0.0010  & 358.8$-$00.0 & Te 2022     & ~~~0.0006
\nl
039.5$-$02.7 & M 2-47    & ~~~0.0006  & 359.2$+$01.2 & 19W32       & ~~~0.005 
\nl
040.4$-$03.1 & K 3-30    & ~~~0.0019  & 359.3$+$01.4 & Th 3-35     & ~~~0.006 
\nl
053.8$-$03.0 & A 63      & ~~~0.0020  & 359.3$-$00.9 & Hb 5        & ~~~0.005 
\nl
068.7$+$01.9 & K 4-41    & ~~~0.0020  & 359.8$+$02.4 & Th 3-33     & ~~~0.003 
\nl
068.7$+$03.0 & PC 23     & ~~~0.0023  & 359.8$-$07.2 & M 2-32      & ~~~0.003 
\nl
076.3$+$01.1 & A 69      & ~~~0.0012  &            &             &       \nl
\enddata
\end{deluxetable}
 
\begin{deluxetable}{llrrrrrrr}
\tablenum{5}
\tablewidth{44pc}
\tablecaption{Planetary Nebulae Parameters}
\tablehead{
\multicolumn{1}{l}{PN G Name} & 
\multicolumn{1}{l}{PN Name} & 
\multicolumn{1}{c}{$d$} & 
\multicolumn{1}{c}{$A_{\rm V}$} &
\multicolumn{1}{c}{$N_{\rm H}$} & 
\multicolumn{1}{c}{$\theta$} & 
\multicolumn{1}{c}{$N_{\rm e}$} & 
\multicolumn{1}{c}{$T_{\rm eff}$} &
\multicolumn{1}{c}{$\log{(g)}$} \\
\multicolumn{1}{c}{$l~~~~~~~b~$} & 
\multicolumn{1}{c}{} & 
\multicolumn{1}{c}{[pc]} &
\multicolumn{1}{c}{[mag]} & 
\multicolumn{1}{c}{[10$^{21}$ cm$^{-2}$]} & 
\multicolumn{1}{c}{[\arcsec]} &
\multicolumn{1}{c}{[cm$^{-3}$]} & 
\multicolumn{1}{c}{[K]} & 
\multicolumn{1}{c}{} 
}

\startdata 
000.1$-$01.1 & M\,3$-$43             &    5819 &     7.8~ &    14.4~~~~~ &    
1.9~~ &	  1550 & \nodata & $\dots$~~ \nl
000.1$+$02.6 & Al\,2$-$J             & \nodata & $\dots$~ & $\dots$~~~~~ &
$\dots$~~ & \nodata & \nodata & $\dots$~~ \nl
000.5$-$01.6 & Al\,2$-$Q             & \nodata & $\dots$~ & $\dots$~~~~~ &
$\dots$~~ & \nodata & \nodata & $\dots$~~ \nl
000.6$-$01.3 & Bl\,3$-$15            & \nodata &     8.3~ &    15.2~~~~~ &
$\dots$~~ & \nodata & \nodata & $\dots$~~ \nl
000.7$-$03.7 & M\,3$-$22             &    7641 &     2.1~ &	3.8~~~~~ &    
3.2~~ & \nodata &  104500 &     5.5~~ \nl
000.8$-$01.5 & Bl\,O                 & \nodata &     4.0~ &	7.4~~~~~ &
$\dots$~~ & \nodata & \nodata & $\dots$~~ \nl
001.2$-$03.9 & ShWi\,2$-$5           & \nodata &     2.5~ &	4.6~~~~~ &
$\dots$~~ & \nodata &   90000\tablenotemark{1} & $\dots$~~ \nl
002.2$+$00.5 & Te\,2337              & \nodata & $\dots$~ & $\dots$~~~~~ &
$\dots$~~ & \nodata & \nodata & $\dots$~~ \nl
002.6$-$03.4 & M\,1$-$37             &    5830 &     2.5~ &	4.6~~~~~ &    
1.4~~ &    4000 &   37800 &     3.7~~ \nl
003.3$-$07.5 & KFL\,19               & \nodata &     1.1~ &	2.0~~~~~ &
$\dots$~~ & \nodata & \nodata & $\dots$~~ \nl
003.5$-$04.6 & NGC\,6565             &    4616 &     0.9~ &	1.7~~~~~ &    
4.5~~ &    1460 &  144000 &     7.3~~ \nl
003.8$-$04.3 & H\,1$-$59             &    9348 &     2.2~ &	4.0~~~~~ &    
3.0~~ &    1350 & \nodata & $\dots$~~ \nl
007.8$-$03.7 & M\,2$-$34             & \nodata &     3.7~ &	6.8~~~~~ &
$\dots$~~ & \nodata & \nodata & $\dots$~~ \nl
009.8$-$07.5 & GJJC\,1               & \nodata & $\dots$~ & $\dots$~~~~~ &
$\dots$~~ & \nodata & \nodata & $\dots$~~ \nl
019.9$+$00.9 & M\,3$-$53             & \nodata &     6.9~ &    12.8~~~~~ &
$\dots$~~ & \nodata & \nodata & $\dots$~~ \nl
021.7$-$00.6 & M\,3$-$55             & \nodata &     5.2~ &	9.6~~~~~ &    
3.6~~ & \nodata & \nodata & $\dots$~~ \nl
021.8$-$00.4 & M\,3$-$28             &    4856 &     6.3~ &    11.6~~~~~ &    
4.5~~ & \nodata &  130500 &     5.8~~ \nl
031.9$-$00.3 & WeSb\,4               & \nodata & $\dots$~ & $\dots$~~~~~ &
$\dots$~~ & \nodata & \nodata & $\dots$~~ \nl
034.6$+$11.8 & NGC\,6572             &     705 &     0.7~ &	1.4~~~~~ &    
7.2~~ &	  4960 &   56900\tablenotemark{2} & $\dots$~~ \nl
036.1$-$57.1 & NGC\,7293             &     157 &    0.09~ &    0.16~~~~~ &  
402.0~~ &	   160 &  100000 &     6.9~~ \nl
037.7$-$34.5 & NGC\,7009             &    1201 &    0.24~ &    0.44~~~~~ &   
14.1~~ &	  3860 &   82000\tablenotemark{3} &     4.8~~ \nl
039.5$-$02.7 & M\,2$-$47             &    4743 &     3.8~ &	7.0~~~~~ &    
4.1~~ &	  2750 & \nodata & $\dots$~~ \nl
040.4$-$03.1 & K\,3$-$30             &    6291 &     4.0~ &	7.3~~~~~ &    
1.7~~ & \nodata & \nodata & $\dots$~~ \nl
053.8$-$03.0 & A\,63                 & \nodata &     2.6~ &	4.8~~~~~ &
$\dots$~~ & \nodata & \nodata & $\dots$~~ \nl
060.8$-$03.6 & NGC\,6853             &     262 &     0.4~ &	0.7~~~~~ &  
170.0~~ &	   250 &  110100 &     7.3~~ \nl
064.7$+$05.0 & BD$+$30$^{\circ}$3639 &    1162 &     1.0~ &	1.8~~~~~ &    
2.4~~ &   10000 &   31100 &     3.1~~ \nl
068.7$+$01.9 & K\,4$-$41             &    7929 &     3.5~ &	6.4~~~~~ &    
1.5~~ & \nodata & \nodata & $\dots$~~ \nl
068.7$+$03.0 & PC\,23                &    6680 &     3.7~ &	6.8~~~~~ &    
1.1~~ & \nodata & \nodata & $\dots$~~ \nl
076.3$+$01.1 & A\,69                 &    4173 &     4.9~ &	9.0~~~~~ &   
11.0~~ & \nodata & \nodata & $\dots$~~ \nl
077.5$+$03.7 & KjPn\,1               & \nodata &     3.7~ &	6.8~~~~~ &
$\dots$~~ & \nodata & \nodata & $\dots$~~ \nl
077.7$+$03.1 & KjPn\,2               & \nodata & $\dots$~ & $\dots$~~~~~ &
$\dots$~~ & \nodata & \nodata & $\dots$~~ \nl
088.7$+$04.6 & K\,3$-$78             &    7831 & $\dots$~ & $\dots$~~~~~ &    
2.2~~ & \nodata & \nodata & $\dots$~~ \nl
094.0$+$27.4 & K\,1$-$16             &    1003 &     0.0~ &	0.0~~~~~ &   
47.0~~ &   102\tablenotemark{4} &  138000\tablenotemark{5} &     6.1~~ \nl
096.4$+$29.9 & NGC\,6543             &     982 &     0.3~ &	0.5~~~~~ &    
9.4~~ &    2700 &   47000 &     3.8~~ \nl
104.2$-$29.6 & Jn\,1                 &     709 &     0.6~ &	1.1~~~~~ &  
166.0~~ &	 15\tablenotemark{4} &   95000 & $\dots$~~ \nl
110.6$-$12.9 & K\,1$-$20             &    4178 &     1.1~ &	2.0~~~~~ &   
17.0~~ & \nodata & \nodata & $\dots$~~ \nl
118.8$-$74.7 & NGC\,246              &     629 &     0.3~ &	0.5~~~~~ &  
112.0~~ &   200\tablenotemark{4} &   85000\tablenotemark{3} &     5.7~~ \nl
130.4$+$03.1 & K\,3$-$92             &    6806 &     3.0~ &	5.5~~~~~ &    
6.5~~ &	   130 &  115000 & $\dots$~~ \nl
148.4$+$57.0 & NGC\,3587             &     615 &    0.02~ &    0.04~~~~~ &  
100.0~~ &     125 &  104000 & $\dots$~~ \nl
183.8$+$05.5 & WeSb\,2               & \nodata & $\dots$~ & $\dots$~~~~~ &
$\dots$~~ & \nodata & \nodata & $\dots$~~ \nl
189.1$+$19.8 & NGC\,2371$-$72        &    1539 &     0.3~ &	0.5~~~~~ &   
21.8~~ &    1700 &   68300 &     5.1~~ \nl
197.8$+$17.3 & NGC\,2392             &    1149 &     0.5~ &	0.9~~~~~ &   
22.4~~ &	  1280 &   47000\tablenotemark{6} &     3.8~~ \nl
205.1$+$14.2 & A\,21                 &     243 &     0.0~ &	0.0~~~~~ &  
307.5~~ &	   180 &  137000 & $\dots$~~ \nl
206.4$-$40.5 & NGC\,1535             &    2283 &    0.22~ &    0.40~~~~~ &    
9.2~~ &	  2140 &  138000\tablenotemark{3} &     6.5~~ \nl
208.5$+$33.2 & A\,30                 &    1689 &     0.0~ &	0.0~~~~~ &   
63.5~~ &	    20 &  116000 &     5.5~~ \nl
215.2$-$24.2 & IC\,418               &     609 &     0.7~ &	1.2~~~~~ &    
6.2~~ &	  4100 &   36000 &     3.5~~ \nl
219.1$+$32.2 & A\,31                 &     233 &     0.0~ &	0.0~~~~~ &  
486.0~~ &	  10\tablenotemark{4} &  147000 & $\dots$~~ \nl
220.3$-$53.9 & NGC\,1360             &     350 &     0.0~ &	0.0~~~~~ &  
192.0~~ &	   100 &   80000\tablenotemark{3} &     5.4~~ \nl
233.5$-$16.3 & A\,15                 &    3656 &    0.22~ &	0.4~~~~~ &   
17.0~~ &	    50 &   65000\tablenotemark{3} &     4.1~~ \nl
238.0$+$34.8 & A\,33                 &     751 &     0.8~ &	1.5~~~~~ &  
134.0~~ &	30\tablenotemark{4} &	86000\tablenotemark{1} & $\dots$~~ \nl
285.4$+$01.5 & Pe\,1$-$1             &    2219 &     3.9~ &	7.2~~~~~ &    
1.5~~ &   10940 &   78000\tablenotemark{7} & $\dots$~~ \nl
286.8$-$29.5 & K\,1$-$27             &    1300 &    0.15~ &    0.28~~~~~ &   
53.0~~ &	  50\tablenotemark{8} &  110000 &     6.5~~ \nl
318.4$+$41.4 & A\,36                 &     243 &    0.22~ &    0.40~~~~~ &  
183.5~~ &	36\tablenotemark{4} &	95000\tablenotemark{6} &     5.3~~ \nl
326.1$-$01.9 & VBe\,3                & \nodata &     2.6~ &	4.8~~~~~ &
$\dots$~~ & \nodata & \nodata & $\dots$~~ \nl
339.9$+$88.4 & LoTr\,5               &     616 &     0.0~ &    0.00~~~~~ &    
510~~ &      20 &  150000 & $\dots$~~ \nl
346.3$-$06.8 & Fg\,2                 & \nodata &     1.1~ &	2.1~~~~~ &
$\dots$~~ & \nodata & \nodata & $\dots$~~ \nl
352.9$-$07.5 & Fg\,3                 &    2249 &     1.5~ &	2.8~~~~~ &    
1.0~~ & \nodata &   49000\tablenotemark{7} & $\dots$~~ \nl
358.8$-$00.0 & Te\,2022              & \nodata & $\dots$~ & $\dots$~~~~~ &
$\dots$~~ & \nodata & \nodata & $\dots$~~ \nl
359.2$+$01.2 & 19W32    & \nodata & 6.3\tablenotemark{9}~ &    11.6~~~~~ &   
23.0~~ & \nodata & \nodata & $\dots$~~ \nl
359.3$+$01.4 & Th\,3$-$35            & \nodata &     6.5~ &    12.0~~~~~ &
$\dots$~~ & \nodata & \nodata & $\dots$~~ \nl
359.3$-$00.9 & Hb\,5                 &    1242 &     4.1~ &	7.5~~~~~ &   
10.0~~ &    6280 &  131000 &     5.9~~ \nl
359.8$+$02.4 & Th\,3$-$33            & \nodata &     7.6~ &    14.0~~~~~ &
$\dots$~~ & \nodata & \nodata & $\dots$~~ \nl
359.8$-$07.2 & M\,2$-$32             & \nodata &     0.9~ &	1.6~~~~~ &
$\dots$~~ & \nodata &   60000\tablenotemark{7} & $\dots$~~ \nl
\enddata

\tablerefs{All data come from Cahn, Kaler, \& Stanghellini (1992) unless 
otherwise marked with one of the following superscripts: 
(1) Shaw \& Kaler (1989); (2) de Freitas et al (1986); 
(3) M\'endez et al.\ (1988); (4) Phillips (1988); (5) Kaler (1983); 
(6) McCarthy et al.\ (1990); (7) Gleizes, Acker, \& Stenholm (1989); 
(8) Rauch, K\"oppen, \& Werner (1994); and (9) Corradi (1995).}

\end{deluxetable}

\clearpage

\clearpage

\begin{figure}
\includegraphics{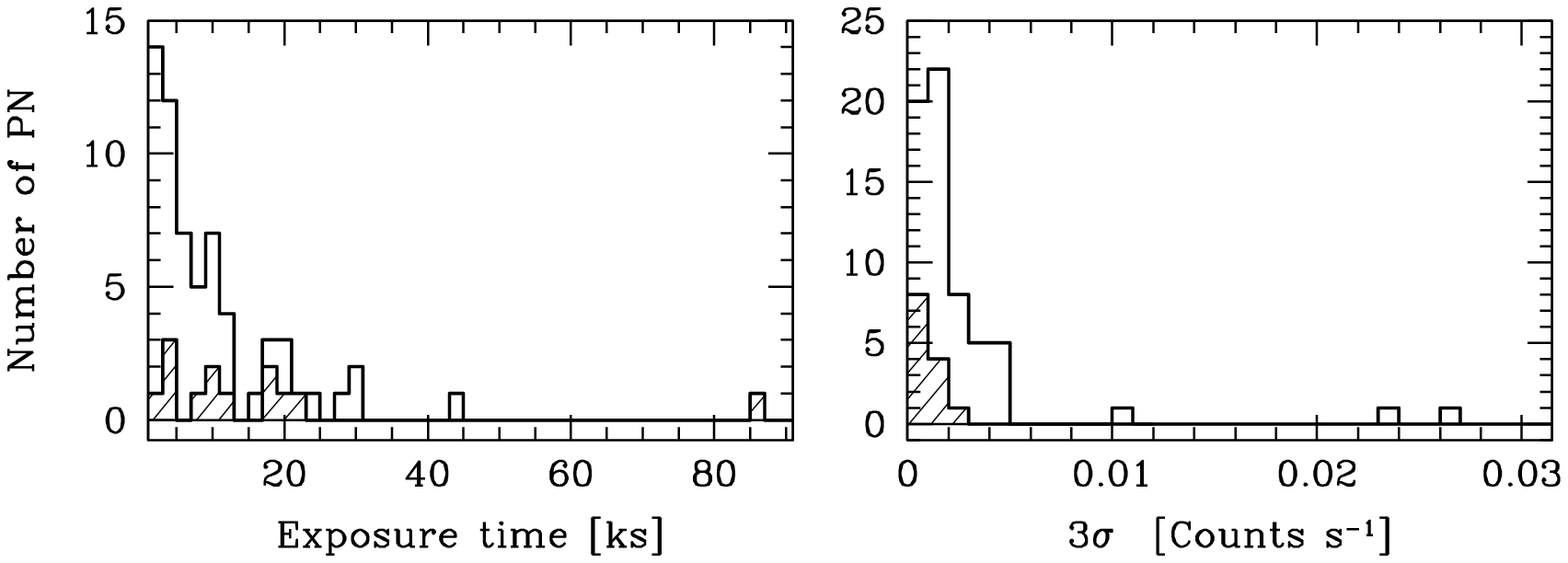}
\end{figure}
     
{\small
\vspace*{15.0cm}
{\sc Fig.~1}--
Observed distributions of the exposure time [{\it left}] and 3$\sigma$
detection limit [{\it right}] measured for the PSPC observations of the
ROSAT PN sample.
The sample size is 63 PNs, and the number of detections is 13 PNs.
The shaded portion of the histograms show the distribution of detected
PNs.}

\clearpage

%
{\sc Fig.~2}{\it a}-- 
X-ray [{\it left}] and Digital Sky Survey [{\it right}] images of NGC\,7293, 
NGC\,7009, and NGC\,6853. 
Contours of the X-ray emission are overlaid on both images.  
The contour levels are 20, 50, 75, and 90\% of the peak surface brightness 
of the PN X-ray emission.
\\

%
%
{\sc Fig.~2}{\it b}-- 
Same as Fig.~2{\it a}, but for BD$+$30$^{\circ}$3639, K\,1-16, and NGC\,6543.
\\

%
%
{\sc Fig.~2}{\it c}-- 
Same as Fig.~2{\it a}, but for NGC\,246, NGC\,3587, and A\,12 (previously
misidentified as the source of nearby X-ray emission). 
\\

%
%
{\sc Fig.~2}{\it d}-- 
Same as Fig.~2{\it a}, but for A\,30, NGC\,1360, and A\,33 (previously
misidentified as the source of nearby X-ray emission).
\\

%
%
{\sc Fig.~2}{\it e}-- 
Same as Fig.~2{\it a}, but for K\,1-27, A\,36, and LoTr\,5.

\clearpage

\begin{figure}
\includegraphics{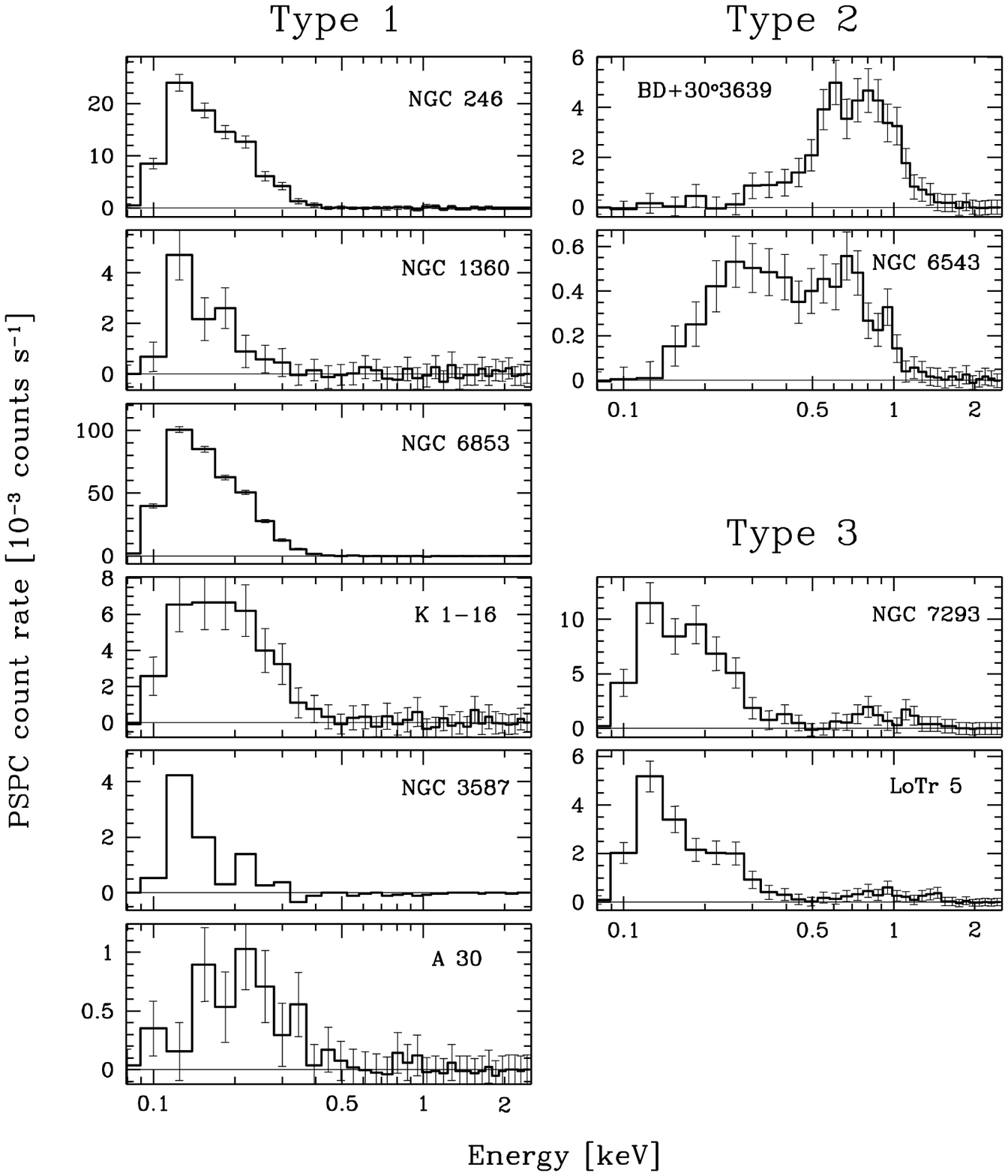}
\end{figure}
     
{\small
\vspace*{21.5cm}
{\sc Fig.~3}--
X-ray spectra for the 10 PNs in the ROSAT PN sample with sufficient S/N
to classify their spectral properties.  The spectra clearly show that
there are three distinct classes of X-ray emission observed in PNs.
}

\clearpage

\begin{figure}
\includegraphics{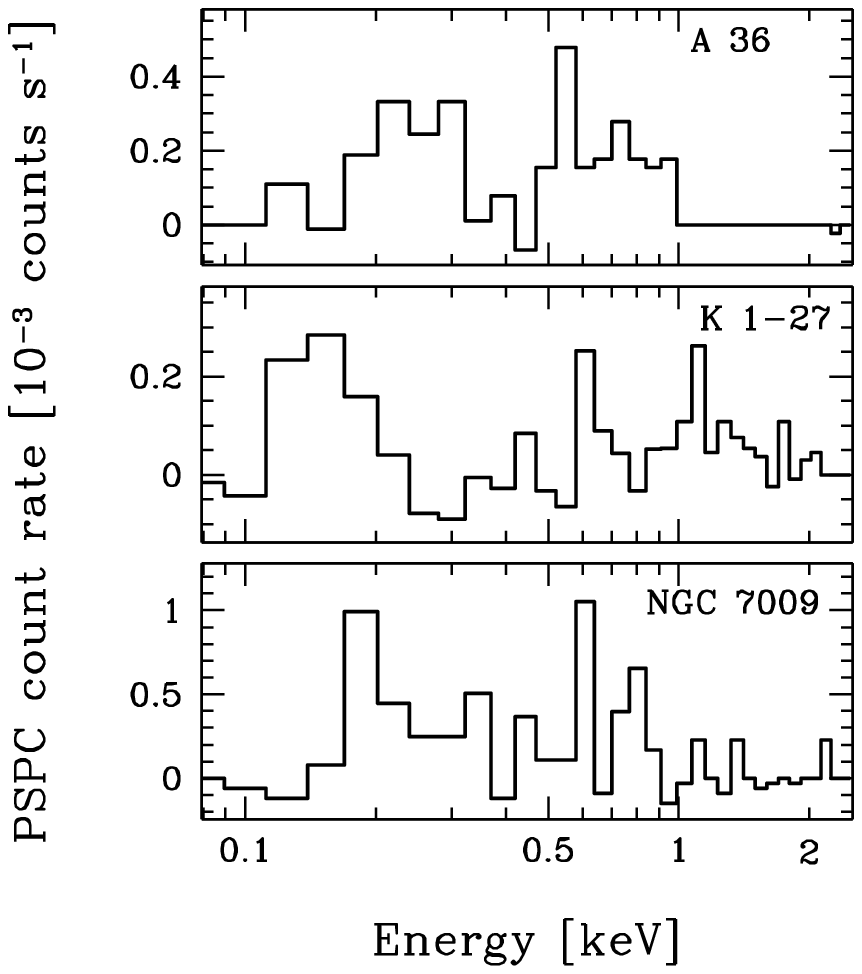}
\includegraphics{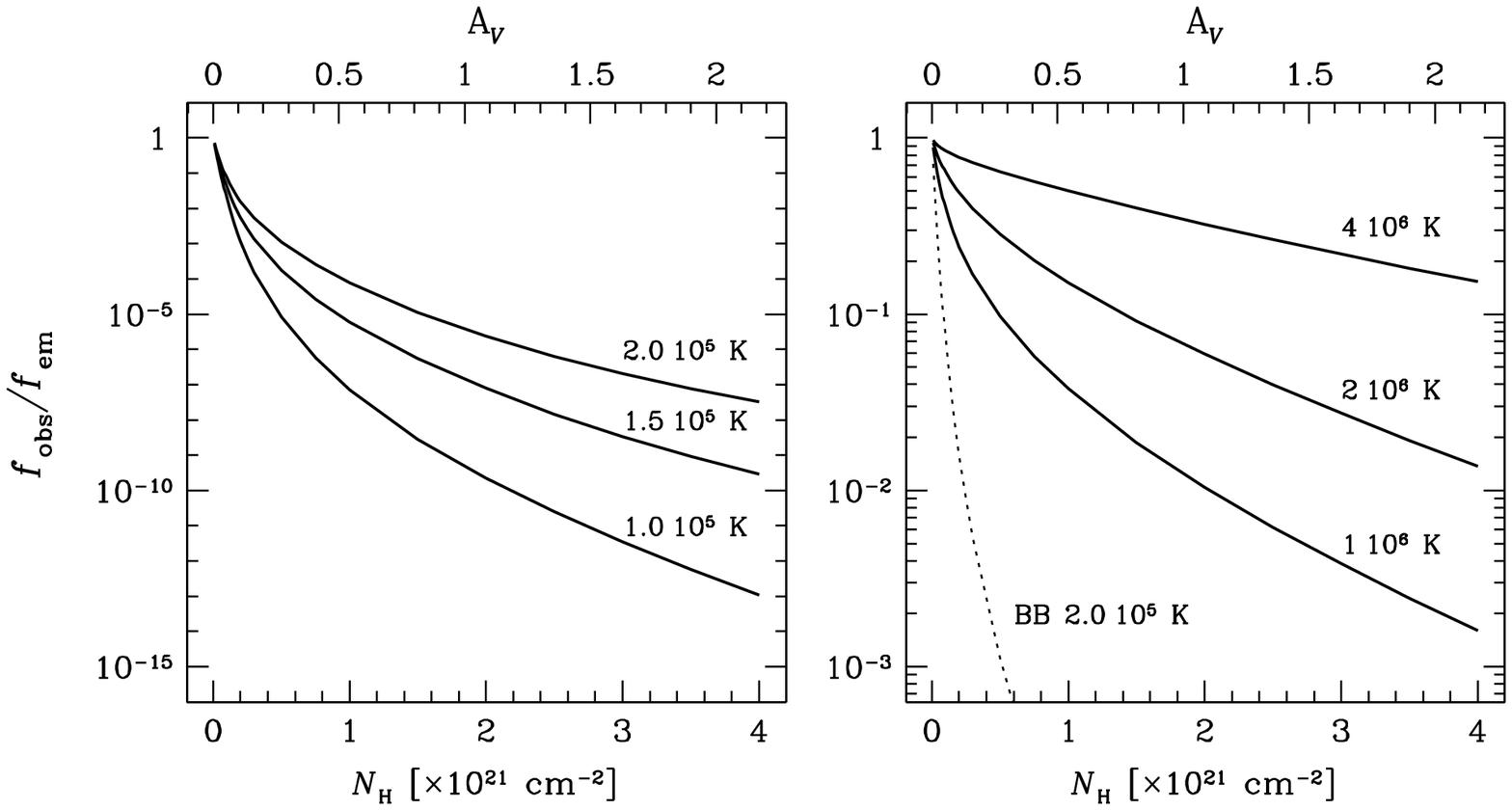}
\end{figure}

{\small
\vspace*{9.5cm}
{\sc Fig.~4}--
X-ray spectra for the 3 PNs in the ROSAT PN sample with insufficient S/N
to classify their spectral properties unambiguously.  The lack of
a prominent peak at 0.1--0.2 keV indicates that these spectra are 
most likely of Type 2.
}
     
{\small
\vspace*{10.5cm}
{\sc Fig.~5}--
Absorbed to unabsorbed flux ratio, $f_{\rm abs}/f_{\rm em}$, in the 
ROSAT PSPC band (0.1--2.4 keV) for a blackbody [{\it left panel}] and 
a thin plasma [{\it right panel}] emission model as a function of the 
absorption column density, $N_{\rm H}$ (or extinction $A_{\rm V}$).  
In order to better show the extreme difference in the detectability of 
emission from the two models, the curve corresponding to the highest 
temperature blackbody model ($T_{\rm eff}=2.0\times10^5$~K) has also been 
included ({\it dotted line}) in the right panel.
}

\clearpage

\begin{figure}
\includegraphics{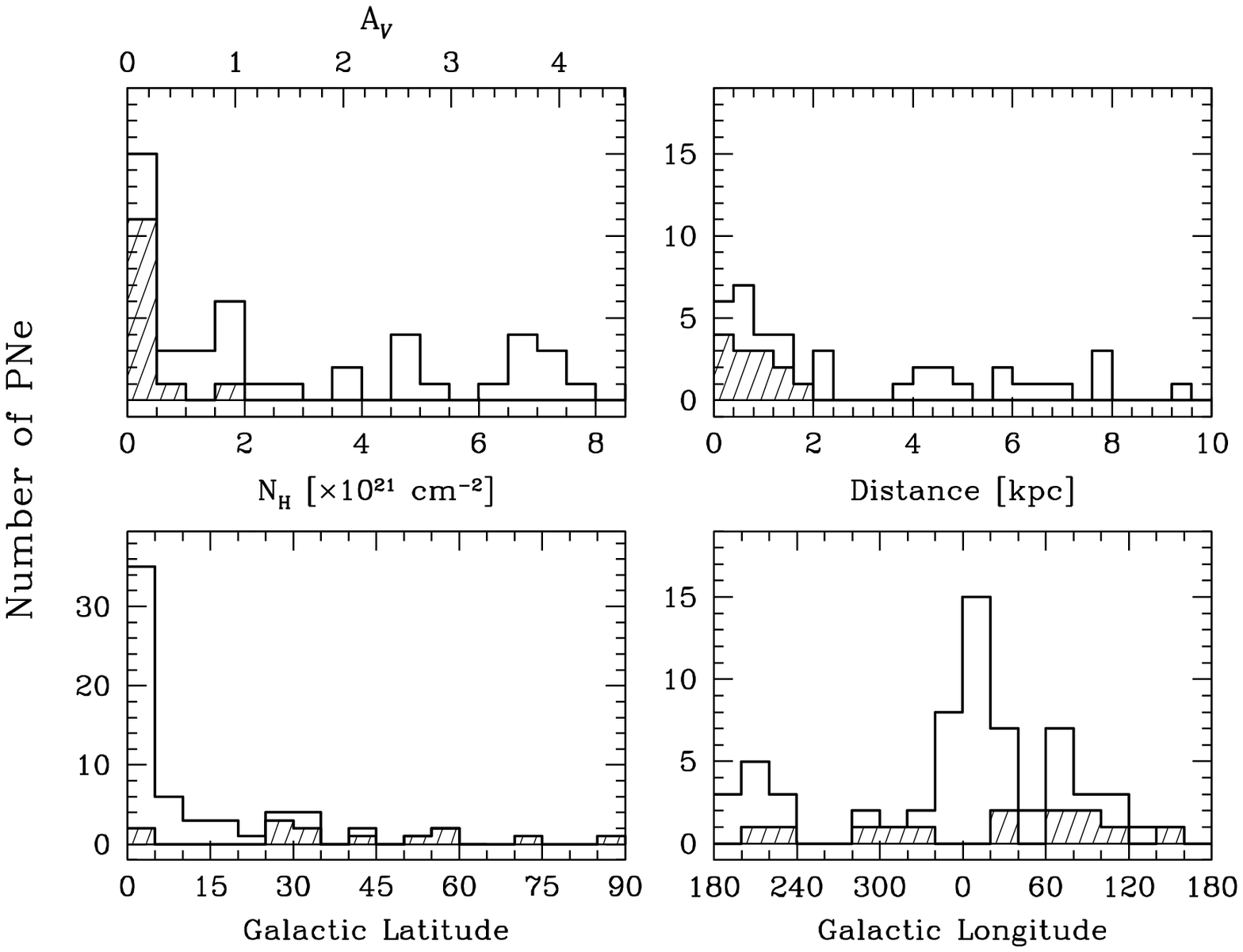}
\includegraphics{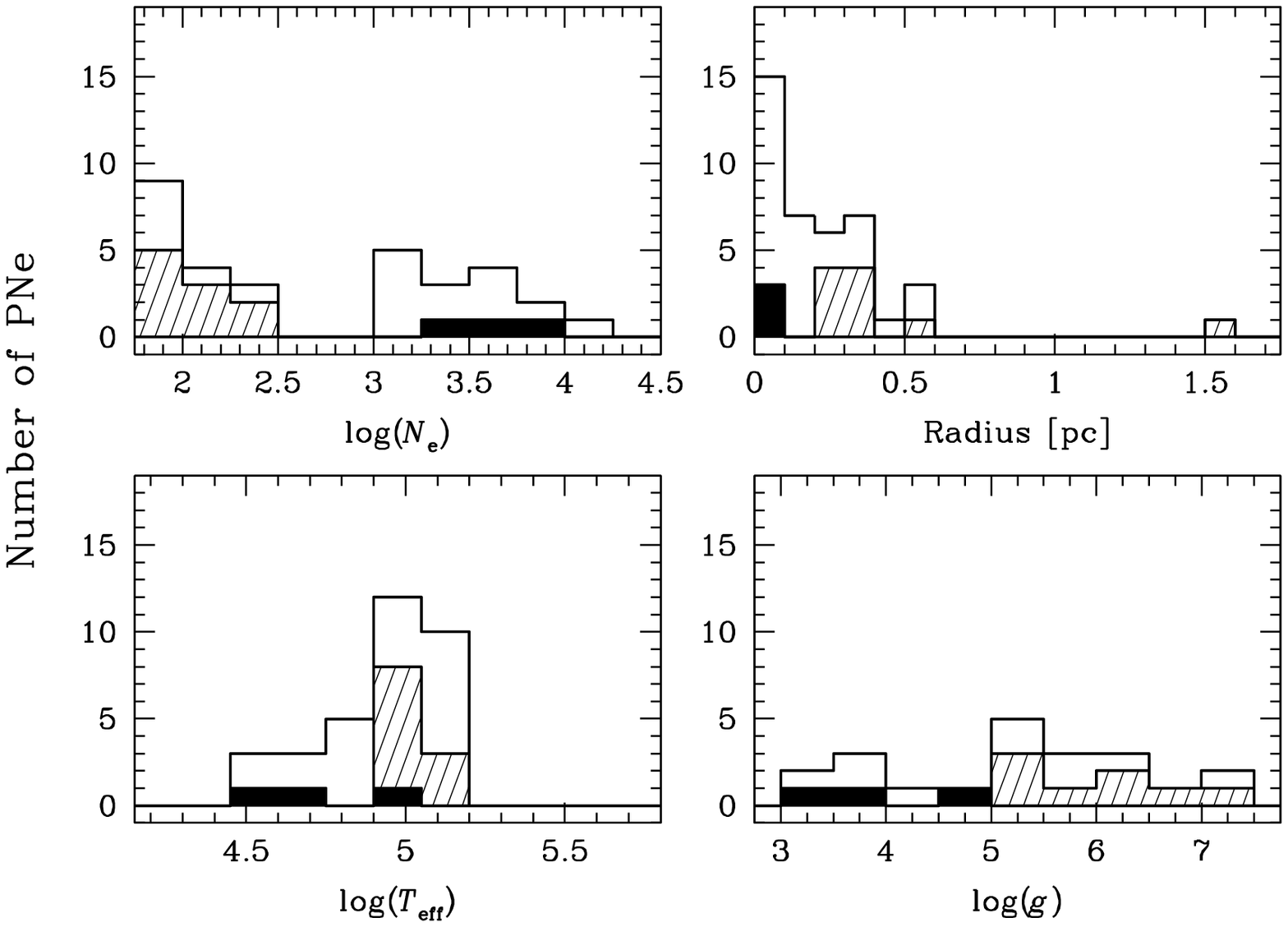}
\end{figure}

{\small
\vspace*{10.0cm}
{\sc Fig.~6}--
Observed distributions of the galactic coordinates, column densities, 
and distances for the ROSAT PN sample.  
As in Figure~1, the shaded histogram shows the distribution
of detected PNs while the open histogram shows the distribution of all
PNs.
}

{\small
\vspace*{10.5cm}
{\sc Fig.~7}--
Observed distributions of the effective temperature and gravity of the 
central stars, and nebular radius and density for the ROSAT PN sample 
(open histogram).
The PNs with detected X-ray emission and Type 1 or 3 spectra are shown 
by the shaded histogram, whereas the PNs with Type 2 spectra are shown 
by the solid histogram.
}

\end{document}